\documentclass[aps,prd,showpacs,preprintnumbers,twocolumn,groupedaddress,nofootinbib]{revtex4-2}
\usepackage{graphicx} 
\usepackage{amsmath}
\usepackage[]{amssymb}
\usepackage[]{mathrsfs}
\usepackage[]{eucal}
\usepackage[]{tensor}
\usepackage[]{amsthm}
\usepackage[]{bm}
\usepackage{color}

\newcommand{\pf}[1]{\mathbf{#1}}

\newcommand{\hdg}{\star} 
\newcommand{\df}{\mathrm{d}}

\newcommand{\bc}{\begin{center}}
\newcommand{\ec}{\end{center}}
\newcommand{\be}{\begin{equation}}
\newcommand{\ee}{\end{equation}}

\newcommand{\FF}{\mathcal{F}}
\newcommand{\GG}{\mathcal{G}}
\newcommand{\LL}{\mathscr{L}}

\begin{document}

\title{\bf Exact multiblack hole spacetimes in Einstein-ModMax theory}
\author{Ana Bokuli\'c}
\email{abokulic.phy@pmf.hr}
\affiliation{Department of Physics, Faculty of Science, University of Zagreb, Bijeni\v cka cesta 32,
10000 Zagreb, Croatia}
\affiliation{Departamento de Matem\'atica da Universidade de Aveiro and Centre for Research and Development  in Mathematics and Applications (CIDMA), Campus de Santiago, 3810-193 Aveiro, Portugal.}
\author{Carlos A. R. Herdeiro}
\email{herdeiro@ua.pt}
\affiliation{Departamento de Matem\'atica da Universidade de Aveiro and Centre for Research and Development  in Mathematics and Applications (CIDMA), Campus de Santiago, 3810-193 Aveiro, Portugal.}

\begin{abstract}
Exact solutions describing multiple, electrically charged black holes (BHs) in a model of nonlinear electrodynamics (NLE) minimally coupled to Einstein's gravity are presented. The NLE model is ModMax theory, that has attracted much attention due to its duality and conformal invariance, features shared with standard (linear) electrodynamics. In the nonextremal case, the solution has conical singularities, similarly to the multi Reissner-Nordstr\"om solution in Einstein-Maxwell theory. In the extremal case the solution is regular on and outside the event horizon; it is isometric to the Majumdar-Papapetrou solution, although the individual BHs have a nonunitary charge to mass ratio, due to screening effects. Using the ModMax electromagnetic duality invariance, magnetically charged and dyonic generalizations are also obtained. Finally, we construct multi-BH solutions with a positive cosmological constant.
\end{abstract}


\maketitle

\section{Introduction}
Observational evidence, especially from gravitational waves, confirms that binary black holes (BHs) are generic astrophysical objects. These can be highly dynamical, and the corresponding solutions of general relativity (GR) have to be obtained numerically. Remarkably, however, GR does allow for \textit{static or stationary} binaries that can be constructed analytically. Even though these have a limited interest for describing astrophysical systems, they can provide insight into strong field, nonlinear features of multicentre compact objects in GR.

The pioneering example of a BH (static) binary in GR is the double Schwarzschild solution \cite{WB22, CP00}, which was subsequently extended to a many (colinear) BH system \cite{IK64}. The backbone of this multi-BH solution is the Weyl formalism \cite{Weyl17}, used to turn nonlinear Einstein equations into a partially linearised system, with the linear part consisting of a Laplace equation. Consequently,  the superposition principle holds, and the addition of more BHs follows a simple algorithmic approach.  Since a single BH is represented by a Newtonian potential of a homogeneous rod \cite{Stephani}, obtaining multiple BHs is reduced to the summation of rod potentials. This tactic, augmented with another algorithmic procedure known as the \textit{inverse scattering method}~\cite{Belinsky:1979mh}, has been successfully applied to construct also Kerr-like double BHs \cite{KN80}, from which one can extract some physical information about their interaction - see $e.g.$~\cite{CHR09}. The Weyl construction can also be applied to obtain  higher-dimensional vacuum multi-BH spacetimes \cite{ER02}.


All of these multi-BH spacetimes, albeit computed as vacuum solutions, actually require localised sources outside the horizons diagnosed from the existence of conical singularities~\cite{Einstein:1936fp}.\footnote{Exceptions exist in higher spacetime dimensions, linked to the existence of topologically nontrivial horizons.} Such singularities are a consequence of unbalanced gravitational interaction. When positioned between BHs, they may be interpreted as struts that are opposing the gravitational attraction. Another option is to depict them as strings extending from the horizons  to infinity. Both interpretations have their shortcomings in terms of physical plausibility. 

There are, however, known multi-BH solutions without such pathologies.  Their existence is based on two main ingredients: introducing an extra repulsive interaction and fine-tuning of the solution parameters. One possibility is the addition of the scalar hair to the two-BH system \cite{HR23D,HR23}. Conical singularities may also be evaded by immersing multiple BHs into an external environment.  These are nonasymptotically flat examples, and examples known in closed form include, $e.g.$ multi-BHs in a Melvin universe \cite{EMP00} or in a de Sitter background \cite{KT93}. Another obvious candidate for modeling a repulsive force is the electromagnetic interaction.  In their seminal papers, Majumdar \cite{Majumdar47} and Papapetrou \cite{Papa47} (independently) showed that a static equilibrium state is achieved if all charged BHs are made extremal. Moreover, the BHs can be placed at arbitrary relative positions. In a way this is in agreement with a Newtonian intuition: the Newtonian gravitational attraction and Coulombic electrostatic repulsion have the same functional form, and therefore can be exactly balanced between two charged point masses at any given distance, if the charge to mass ratio is unity, in appropriate units. This naive interpretation, however, raises an intriguing question: can a modification of classical electromagnetism provide a similar balancing effect, and if so, under what conditions? 

Various generalisations of Maxwell's theory are joined under the umbrella of \textit{nonlinear electrodynamics} (NLE), usually motivated to describe high-energy phenomena in electromagnetic interactions, a regime beyond the validity of the (classical) Maxwell theory. Amongst the first formulated NLE theories were the Born-Infeld electrodynamics \cite{Born34, BI34} and the Euler-Heisenberg \cite{HE36} models. The latter has its roots in quantum field theory as it describes (effectively) one-loop QED effects in two-photon interaction, while the former cures singularities in the electrostatic quantities associated with point charges. Within the landscape of NLE models, a recently introduced one, dubbed \textit{ModMax theory}~\cite{BLST20}, has got considerable attention, partly because it bears similarity with Maxwell's electrodynamics in its underlying symmetries. In fact, it is a unique one-parameter theory that is invariant under $SO(2)$ duality rotations and respects conformal invariance.

When minimally coupled to GR, ModMax electrodynamics - as Maxwell electrodynamics - will certainly yield a rich variety of spacetime solutions, which have only started to be explored. Spherically symmetric, electrically charged ModMax BHs derived in \cite{FAGMLM20} exhibit minimal deviation from their Einstein-Maxwell counterparts, manifesting solely as a charge screening factor. The parameters of these solutions may be constrained by the observational EHT data \cite{PML22}. To analyse the thermodynamic properties of ModMax BHs, one can apply methods presented in \cite{BJS21}. A class of Taub-NUT solutions and dyonic anti-de-Sitter BHs sourced by ModMax electrodynamics were obtained in \cite{BORDO2021136312} and \cite{Flores-Alonso2021}. The accelerating ModMax BH \cite{BARRIENTOS2022137447} serves as the basis for the construction of Melvin-Bonnor-ModMax background \cite{BCPK24}, in which one may embed Schwarzschild and C-metric BHs. Other possible generalisations fall into two main categories: the ones with additional matter degrees of freedom besides ModMax electromagnetic fields and the ones within modified gravitational theories. An example of the former is the NUT- like solution with conformally coupled scalar field \cite{PhysRevD.104.084094, PhysRevD.110.064027}, while the latter includes spherically symmetric ModMax BHs in Einstein-Gauss-Bonnet gravity \cite{ALI2022168726} and $f(R)$ theory \cite{10.1093/ptep/ptae012}. Even though the family of exact, single ModMax BH solutions has expanded, multiple BH systems in ModMax theory remain unexplored, as in fact for generic Einstein-NLE models. In this paper we shall show that Einstein-ModMax theory does allow multi-BH spacetimes, in a \textit{closed analytic form}, and analyse the differences and similarities with respect to the Einstein-Maxwell case.

The paper is organised as follows: in the first section, we introduce NLE and derive the Einstein-ModMax system of equations. In the next section, we discuss the Weyl problem in GR and its application in the context of ModMax electrodynamics, for electrically charged solutions. The central result, the spacetime consisting of multiple ModMax BHs and its properties, is presented in Sec. 3 and in Sec. 4 for the generic Majumdar-Papapetrou-type solutions, both static (asymptotically flat) and dynamical (asympotically de Sitter). Lastly, in Sec. 6 we take advantage of the electromagnetic duality invariance of the ModMax model to generate dyonic and magnetically charged multi-BH solutions. We present our conclusions in Sec. 7. 

\textit{Conventions and notation. }We will use the ``mostly plus" metric convention, $(-,+,+,+)$, and the system of units such that $G=c=4\pi\epsilon_0=1$. Differential forms are written either in index notation or denoted by bold letters. Partial derivatives of the Lagrangian density $\LL$ are abbreviated as $\LL_X=\partial_X\LL$.

\section{Einstein-ModMax equations}
First, we lay out the fundamental elements of NLE theories.  The NLE Lagrangian density $\LL$ is assumed to be a function of the two quadratic  scalar invariants built from the Maxwell tensor $F_{ab}$,  
\begin{align}
\FF=F_{ab}F^{ab}\ \  \text{and}\ \ \GG=F_{ab}\,{\hdg F^{ab}} \ ,
\end{align}
where $\star$ denotes the Hodge dual. 
We will restrict our attention to the minimal coupling of the gravitational and electromagnetic action, so the total Lagrangian 4-form is just a sum of the Einstein-Hilbert term and NLE contribution,
\begin{align}\label{eq:Lagr}
\pf{L}=\dfrac{1}{16\pi}(R+4\LL(\FF, \GG))\bm{\epsilon}  \ ,
\end{align}
where $R$ is the Ricci scalar and $\bm{\epsilon}$ is the volume form. 
The gravitational field equation that follows from (\ref{eq:Lagr}) is
\begin{align}\label{eq:grav}
\tensor{R}{_a_b}-&\dfrac{1}{2}R\tensor{g}{_a_b}=8\pi \tensor{T}{_a_b}\ .
\end{align}
For future discussion, it is convenient to write the NLE energy-momentum tensor as a sum of the ``Maxwell-like" term and the trace term,
\begin{align}\label{eq:EMt}
T_{ab}=-4\LL_\FF &T^{(Max)}_{ab}+\dfrac{1}{4}T g_{ab}\ ,
\end{align}
where Maxwell's energy-momentum tensor and the trace of the NLE energy-momentum tensor are, respectively,  given by
\begin{align}
T^{(Max)}_{ab}&=\dfrac{1}{4\pi}\left(\tensor{F}{_a_c}\tensor{F}{_b^c}-\dfrac{1}{4}g_{ab}\FF\right)\ ,\\
T&=\dfrac{1}{\pi}(\LL-\LL_\FF\FF-\LL_\GG\GG)\ .
\end{align}
The generalised Maxwell's equations may be compactly written as 
\begin{align}\label{eq:gMax}
\df\pf{F}=0\ \text{and}\ \df{\hdg\pf{Z}}=0\ ,
\end{align}
where we introduced the auxiliary 2-form $\pf{Z}$, 
\begin{align}
\textbf{Z}=-4(\LL_\FF \textbf{F}+\LL_\GG{\hdg\textbf{F}})\ .
\end{align}
The electric and magnetic charges are given by the corresponding Komar integrals evaluated over a closed 2-surface $\mathcal{S}$
\begin{align}\label{eq:ch}
Q=\dfrac{1}{4\pi}\oint_\mathcal{S} {\hdg\pf{Z}}\ \  \text{and}\ \  P=\dfrac{1}{4\pi}\oint_\mathcal{S}\pf{F}.
\end{align}
We turn our focus to ModMax electrodynamics \cite{BLST20, Kosyakov20}, defined by the following Lagrangian:
\begin{align}
\mathcal{L}^{(MM)}=\dfrac{1}{4}\left(-\FF\text{cosh}\gamma+\sqrt{\FF^2+\GG^2}\text{sinh}\gamma\right) . 
\end{align}
To preserve causality, the parameter $\gamma$ should be positive. Notice that for $\gamma=0$ one recovers Maxwell's Lagrangian density.

\section{Weyl formalism for ModMax electrodynamics}
Using the Weyl construction, we will show how the partial linearisation of Einstein's field equations in the vacuum case may be indirectly carried over to Einstein-ModMax theory.  Start from the metric of a static and axisymmetric spacetime, in Weyl form \cite{Weyl17}
\begin{align}
\df s^2=&-e^{2U(\rho, z)}\df t^2+\nonumber\\&+e^{-2U(\rho,z)}\big[e^{2k(\rho, z)}(\df\rho^2+\df z^2)+\rho^2 \df\phi^2\big]\ .
\label{an1}
\end{align}
A simplification of the gravitational-NLE equations (\ref{eq:grav}) and (\ref{eq:gMax}) can be achieved by restricting to configurations with nonvanishing electric field only. Such a scenario corresponds to choosing the gauge potential ansatz as 
\begin{equation}
    \pf{A}=-\chi(\rho, z)\df t \ .
    \label{an2}
\end{equation} 
Then, the electromagnetic invariant $\GG$ vanishes, while $\LL_\FF$ becomes merely a constant factor, $\LL_\FF=-e^{\gamma}/4$. Furthermore, due to the conformal invariance of the theory,  $T=0$ in (\ref{eq:EMt}). Then, the Einstein-ModMax equations are brought down to
\begin{align}
\tensor{R}{_a_b}=8\pi e^{\gamma}T_{ab}^{(Max)},\label{eq:EMM}\\
\df\pf{F}=0\ \text{and}\ \df{\hdg\pf{F}}=0\label{eq:MM}\ ,
\end{align}
thus making the resemblance with Maxwell's theory manifest in the purely electric case.

The field equations (\ref{eq:EMM}) and (\ref{eq:MM}) with the ansatz (\ref{an1}) and (\ref{an2}) give rise to four independent partial differential equations.
The metric function $U(\rho, z)$ and gauge potential function $\chi(\rho, z)$ constitute a nonlinear system,
\begin{align}
\Delta U&=e^{\gamma}e^{-2U}\left[(\partial_\rho \chi)^2+(\partial_z \chi)^2\right], \ \label{eq:U1} \\ \Delta\chi&=2(\partial_z\chi\partial_zU+\partial_\rho\chi\partial_\rho U )\, \label{eq:chi}\ ,
\end{align}
where the Laplace operator acts on the auxiliary Euclidean 3-space $\df s^2=\df \rho^2+dz^2+\rho^2\df\phi^2$. Once the functions $U(\rho, z)$and $\chi(\rho, z)$ are known, the remaining metric function $k(\rho, z)$ can, in principle,  be obtained by integration from
\begin{align}
\partial_zk&=2\rho(\partial_\rho U\partial_z U-e^{\gamma}e^{-2U}\partial_\rho \chi\partial_z \chi) \ ,\label{eq:k1}\\
\partial_\rho k&=\rho\left[(\partial_\rho U)^2-(\partial_z U)^2\right]-\nonumber\\&-e^{\gamma}e^{-2U}\rho\left[(\partial_\rho \chi)^2-(\partial_z \chi)^2\right]\label{eq:k2}\ .
\end{align}

To find a family of solutions of the coupled partial differential equations (\ref{eq:U1}) and (\ref{eq:chi}), that in particular includes a single and multiple-charged BHs, we will use a modified form of a well-known technique employed in the Einstein-Maxwell case \cite{AK94}.  The central idea is to express functions $U(\rho, z)$ and $\chi(\rho, z)$ in terms of a new function $V(\rho, z)$ that will transform the above-mentioned system of equations into a vacuum Weyl problem. The sought substitution can be found using reverse engineering procedure, relying on the fact that the solution spectrum must contain an electrically charged ModMax BH \cite{FAGMLM20}. Using this fact, we may relate functions $U(\rho, z)$ and $\chi(\rho, z)$ via 
\begin{align}
e^{2U(\rho, z)}=1-\dfrac{2e^{\gamma}}{q}\chi(\rho, z)+e^{\gamma}\chi(\rho, z)^2\ .
\end{align}
Following the approach presented in \cite{AK94},  for the ModMax case we set
\begin{align}
\chi(\rho, z)=\dfrac{qme^{-\gamma}}{R+m}\hspace{5mm} \text{and}\hspace{5mm} e^{2U(\rho, z)}&=\dfrac{R^2-d^2}{(R+m)^2}, 
\end{align}
where $m$ is a constant, $d^2=m^2(1-q^2e^{-\gamma})$, while the function $R(\rho, z)$ is conveniently defined as
\begin{align}
R(\rho, z)=d\dfrac{1+e^{2V(\rho, z)}}{1-e^{2V(\rho, z)}}\ .
\end{align}
Then, the mapping to vacuum problem is given by
\begin{align}
&e^{2U(\rho, z)}=\dfrac{4e^{2V(\rho, z)}(1-e^{-\gamma}q^2)}{(1-e^{2V(\rho, z)}+(1+e^{2V(\rho, z)})\sqrt{1-q^2e^{-\gamma}})^2}\label{eq:U}\ ,\\
&\chi(\rho, z)=\dfrac{qe^{-\gamma}(1-e^{2V(\rho, z)})}{1-e^{2V(\rho, z)}+(1+e^{2V(\rho, z)})\sqrt{1-q^2e^{-\gamma}}}\ .\label{eq:pot}
\end{align}
Consequently, the original problem of finding the solutions of the Weyl-ModMax system is reduced to solving vacuum Weyl equations
\begin{eqnarray}
\Delta V=0 \ ,& \\
 \partial_\rho k=\rho\left[(\partial_\rho V)^2-(\partial_z V)^2\right],\ \ &\partial_zk=2\rho\partial_\rho V\partial_z V\ .
\end{eqnarray}
Since $V(\rho, z)$ is a harmonic function, it can be thought of as a Newtonian potential of a mass distribution lying along the $z$-axis.  Once the function $V(\rho, z)$ is specified, the full solution within Einstein-ModMax theory is given by expressions (\ref{eq:U}) and (\ref{eq:pot}), while the function $k(\rho, z)$ remains unaltered compared to the vacuum case.

To illustrate the idea,  we shall apply the described procedure to reconstruct a charged ModMax BH from a known vacuum solution. In Weyl formalism, the Schwarzschild BH is obtained from the Newtonian potential of a homogeneous rod of length  equal to $2\mu$ placed symmetrically along the $z$-axis,
\begin{align}\label{eq:vacSch}
e^{2V(\rho, z)}&=\dfrac{r_1+r_2-2\mu}{r_1+r_2+2\mu}\ ,
\end{align}
where $r^2_i=\rho^2+\zeta_i^2$, $\zeta_1=z+\mu,$ and $\zeta_2=z-\mu$.  The evaluation of the line integrals (\ref{eq:k1}) and (\ref{eq:k2}) gives
\begin{align}
e^{2k(\rho, z)}=\dfrac{(r_1+r_2-2\mu)(r_1+r_2+2\mu)}{4r_1r_2}\ .
\end{align}
Using (\ref{eq:vacSch}), the complete solution in the Einstein-ModMax theory given by (\ref{eq:U}) and (\ref{eq:pot}) is
\begin{align}
e^{2U(\rho, z)}&=\dfrac{2(1-q^2e^{-\gamma})(z^2-\mu^2+\rho^2+r_1r_2)}{(2\mu+\sqrt{1-e^{-\gamma}q^2}(r_1+r_2))^2}\ ,\\
\chi(\rho, z)&=\dfrac{2e^{-\gamma}q\mu}{2\mu+\sqrt{1-e^{-\gamma}q^2}(r_1+r_2)}\ ,
\end{align}
where the parameters $\mu$ and $q$ are interpreted in terms of mass and electric charge of the BH,
\begin{align}
\mu^2=M^2-e^{-\gamma}Q^2, \ q=Q/M\ .
\end{align}
It can be recast into a standard form by switching to spherical coordinates $(r,\theta)$,
\begin{align}
\rho=\sqrt{(r-M)^2-\mu^2}\text{sin}\theta, \ \ z=(r-M)\text{cos}\theta \ .
\end{align} 
Then, we obtain
\begin{align}\label{eq:1MM}
\df s^2&=-f(r)\df t^2+\dfrac{\df r^2}{f(r)}+r^2(\df \theta^2+\text{sin}^2\theta \df\phi^2),\ \nonumber\\ f(r)&=1-\dfrac{2M}{r}+\dfrac{Q^2e^{-\gamma}}{r^2}\ ,\\
\chi(r)&=\dfrac{Qe^{-\gamma}}{r}\ ,
\end{align}
recovering the solution in~\cite{FAGMLM20}, 
therefore confirming the consistency of the approach. In the next section, relying on the ``rod structure" of Weyl formalism, we construct N-ModMax BHs along the symmetry axis.

\section{N-ModMax black holes}
Accommodating multiple BHs in Weyl formalism amounts to superposing rod potentials \cite{IK64},
\begin{align}\label{eq:Umulti}
e^{2V(\rho, z)}&=\dfrac{r_1+r'_1-\mu_1}{r_1+r'_1+\mu_1}...\dfrac{r_N+r'_N-\mu_N}{r_N+r'_N+\mu_N} \ , 
\end{align}
where
\begin{align}
r_i^2&=\rho^2+\zeta_i^2, \ \ {r'_i}^{2}=\rho^2+{\zeta'_i}^2,\ \nonumber \\ \zeta_i&=z-z_i-\mu_i/2, \ \ \zeta'_i=z-z_i+\mu_i/2 \ .
\end{align}
In this parametrisation, it is assumed that the center of the $i$-th rod is at $z=z_i$ and its length is $\mu_i$. The rods are nonoverlapping,  with $z_1<z_2...<z_N$.  For notational simplicity, one may define 
\begin{align}
Y_{ij}&\equiv r_ir_j+\zeta_i\zeta_j+\rho^2,\ \quad \ Y_{i'j}\equiv r'_ir_j+\zeta'_i\zeta_j+\rho^2 \ .
\end{align}
The line integrals (\ref{eq:k1}) and (\ref{eq:k2}) yield
\begin{align}
e^{2k(\rho, z)}&=\prod_{i=1}^{N}\prod_{j=1}^{N}\left(\dfrac{Y_{i'j}Y_{ij'}}{Y_{ij}Y_{i'j'}}\right)^{1/2}\ .
\end{align}
The spacetime containing N-ModMax BHs is obtained by inserting the vacuum solution (\ref{eq:Umulti}) into the metric function (\ref{eq:U}) and gauge potential (\ref{eq:pot}).  
Its total ADM mass may be read off from the asymptotic behaviour of the metric function $e^{2U(\rho, z)}$,
\begin{align}\label{eq:mass}
M=\sum_{i=1}^{N}M_i=\dfrac{1}{2\sqrt{1-q^2e^{-\gamma}}}\sum_{i=1}^{N}\mu_i\ ,
\end{align}
where $M_i$ can be understood as the ADM mass attributed to the $i$-th BH.  The electric charge (\ref{eq:ch}) evaluated over a sphere at infinity is equal to the sum of the individual BH charges $Q_i$,
\begin{align}\label{eq:charge}
Q=\sum_{i=1}^{N}Q_i=\dfrac{q}{2\sqrt{1-q^2e^{-\gamma}}}\sum_{i=1}^{N}\mu_i \ .
\end{align}
Two types of singularities appear in this spacetime, the first of which are curvature singularities shielded by the BH horizons. The second ones are conical singularities along the symmetry axis. 
Between every $i$-th and $(i+1)$-th BH there is an excess angle equal to
\begin{align}
\dfrac{\delta}{2\pi}&=1-e^{-k(\rho=0,z)}=\nonumber\\
&=1-\prod_{j=1}^{i-1}\dfrac{(z_j-z_{i+1})^2-\frac{1}{4}(\mu_j-\mu_{i+1})^2}{(z_j-z_{i+1})^2-\frac{1}{4}(\mu_j+\mu_{i+1})^2}\times\nonumber\\ &\times\prod_{j=i+1}^{N}\dfrac{(z_j-z_{i})^2-\frac{1}{4}(\mu_j-\mu_{i})^2}{(z_j-z_{i})^2-\frac{1}{4}(\mu_j+\mu_{i})^2} \ .
\end{align}

Inspection of the expressions for mass (\ref{eq:mass}) and electric charge (\ref{eq:charge}) reveal one possible way of remedying conical singularities.  Namely, if one takes the limit as $1-q^2e^{-\gamma}\to0$ and $\mu_i\to0$, the mass-to-charge ratio of every single BH remains well-defined and equal to $e^{-\gamma/2}$.  This condition corresponds to the extremal configurations and one recovers a ModMax variant of a Majumdar-Papapetrou spacetime (with colinear BHs)~\cite{Majumdar47, Papa47},
\begin{align}\label{eq:MP}
&\df s^2=-e^{2U(\rho, z)}\df t^2+e^{-2U(\rho,z)}(\df \rho^2+\df z^2+\rho^2 \df \phi^2),\nonumber\\
&e^{2U(\rho,z)}=\left(1+\sum_i\dfrac{M_i}{r_i}\right)^{-2}, \ r_i=\sqrt{\rho^2+(z-z_i)^2}\ .
\end{align}
Since the metrics of the extremal Reissner-Nordstr\"{o}m solution and the extremal ModMax BH are indistinguishable, it is not surprising that the metric (\ref{eq:MP}) stays the same as in the Einstein-Maxwell case.  However, the difference between the two theories becomes apparent from the gauge potential, which in the ModMax case acquires an additional factor of $e^{-\gamma/2}$,
\begin{align}
\textbf{A}=e^{-\gamma/2}e^{U(\rho,z)}\df t\ .
\end{align}
A further distinction is the mass-to-charge ratio needed to sustain the equilibrium configuration without conical singularities. In Maxwell theory, the required ratio is $M_i/Q_i=1$, while for ModMax theory it is less than 1 as a result of the charge screening effect, $M_i/Q_i=e^{-\gamma/2}<1$.

\section{The ModMax multi-extremal BH solution}
\subsection{Asymptotically flat solution}
A more general form of the spacetime describing multiple extremal ModMax BHs is obtained by modifying the original Majumdar-Papapetrou ansatz in Cartesian coordinates \cite{Majumdar47, Papa47},\footnote{With this gauge choice, $A$ does not vanish at infinity, which is just a matter of convention.}
\begin{align}
\df s^2&=-\dfrac{1}{\Omega(x,y,z)^2}\df t^2+\Omega^{2}(\df x^2+\df y^2+\df z^2),\nonumber \\  \textbf{A}&=\dfrac{e^{-\gamma/2}}{\Omega(x, y, z)}\df t  \ .
\end{align}  
With such a choice,  gravitational field equations and the generalised Maxwell's equation are fully linearised and reduced to a single harmonic equation,
\begin{equation}
    \Delta \Omega(x, y, z)=0 \ ,
\end{equation} 
where the Laplace operator $\Delta$ is defined on the auxiliary Euclidean space $\df s^2=\df x^2+\df y^2+\df z^2$.  One of its possible solutions is
\begin{align}
\Omega=1+\sum_{i=1}^{N}\dfrac{M_i}{\sqrt{(x-x_i)^2+(y-y_i)^2+(z-z_i)^2}} \ ,
\end{align}
corresponding to N-extremal ModMax BHs which are no longer spread out along the $z$-axis, but are placed at the arbitrary positions $(x_i, y_i, z_i)$.
\subsection{Asymptotically de Sitter solution}
Building on a previous result describing Majumdar-Papapetrou spacetime embedded in a universe with a positive cosmological constant \cite{KT93}, we can derive an analogous solution for the ModMax case. Taking the following ansatz for the metric and the gauge potential,
\begin{align}\label{eq:LMP}
\df s^2&=-\dfrac{1}{\Omega(x,y,z)^2}\df t^2+a(t)^2\Omega^{2}(\df x^2+\df y^2+\df z^2),\nonumber\\  \textbf{A}&=\dfrac{e^{-\gamma/2}}{\Omega(x, y, z)}\df t \ \nonumber ,\\
\Omega&=1+\sum_{i=1}^{N}\dfrac{M_i}{a(t)\sqrt{(x-x_i)^2+(y-y_i)^2+(z-z_i)^2}}\ ,
\end{align}
the Einstein-ModMax equations are satisfied if the scale factor $a(t)$ is given by
\begin{align}
a(t)&=e^{Ht}, \ \  H=\pm\sqrt{\dfrac{\Lambda}{3}}\ ,
\end{align}
where $\Lambda$ denotes the cosmological constant. The properties of the solution (\ref{eq:LMP}) could be analysed in the same manner as in \cite{PhysRevD.52.796, PhysRevD.49.840} with the expectation that the conclusions directly translate to the ModMax BHs, provided that the extremality condition has been redefined accordingly.

\section{Dyonic solution}
All of the above considerations were done for purely electrically charged solutions, which allowed us to simplify the equations of motion. However, electromagnetic duality invariance of ModMax theory may be used to generate magnetic and dyonic solutions from the electric ones without further solving the equations of motion explicitly, as we now exemplify.

For a general NLE theory, the $SO(2)$ electromagnetic rotation by an angle $\alpha$ is of the form
\begin{align}
\widetilde{\textbf{F}}=\text{sin}\alpha\,{\hdg \textbf{Z}}+\text{cos}\alpha\, \textbf{F}\ ,\label{eq:so(2)1}\\
\widetilde{\textbf{Z}}=\text{cos}\alpha \,\textbf{Z}+\text{sin}\alpha\,{\hdg \textbf{F}}\ \label{eq:so(2)2}.
\end{align}
If the theory is duality invariant, the transformations (\ref{eq:so(2)1}) and (\ref{eq:so(2)2}) can be interpreted as mappings between two different electromagnetic configurations, with the parameter $\alpha$ defining the orbit of solutions.
Taking the electrically charged ModMax solution as a seed, the $SO(2)$ electromagnetic rotations (\ref{eq:so(2)1}) and (\ref{eq:so(2)2}) become
\begin{align}
\widetilde{\textbf{F}}=\text{sin}\alpha\, e^{\gamma}{\hdg \textbf{F}}+\text{cos}\alpha \textbf{F}\ ,\\
\widetilde{\textbf{Z}}=\text{cos}\alpha \,e^{\gamma}\textbf{F}+\text{sin}\alpha{\hdg \textbf{F}}\ .
\end{align}
Using Eqs. (\ref{eq:gMax}), it is straightforward to show that the generalised Maxwell's equations are invariant under the above-mentioned  transformations, 
\begin{align}
\df{\hdg \widetilde{\textbf{Z}}}=0 \ \ \text{and}\ \  \df\widetilde{\textbf{F}}=0\ .
\end{align}
Komar integrals imply that 
\begin{equation}
    \widetilde{Q}=Q\,\text{cos}\alpha \qquad  {\rm and} \qquad \widetilde{P}=Q\,\text{sin}\alpha \ ,
\end{equation} 
$i.e.$ the original charge $Q$ is split into the electric and magnetic parts. In the gravitational sector, it remains to prove a correspondence between the electrically charged solutions of Einstein-ModMax equations defined by $[g_{ab}, F_{ab}]$ and solutions given by $[g_{ab}, \widetilde{F}_{ab}]$. A simple calculation shows that the energy-momentum tensor is unaffected by the duality rotation,
\begin{align}
&\widetilde{T}_{ab}=-4\widetilde{\LL}_\FF \widetilde{T}^{(Max)}_{ab}=\dfrac{e^{\gamma}}{4\pi(\text{cos}^2\alpha+e^{2\gamma}\text{sin}^2\alpha)}\times\nonumber\\
&\times\Big((e^{\gamma}\text{sin}\alpha {\hdg\tensor{F}{_a_c}}+\text{cos}\alpha\tensor{F}{_a_c})(e^{\gamma}\text{sin}\alpha{\hdg\tensor{F}{_b^c}}+\text{cos}\alpha\tensor{F}{_b^c})-\nonumber\\
&-\dfrac{1}{4}\tensor{g}{_a_b}\FF(\text{cos}^2\alpha-e^{2\gamma}\text{sin}^2\alpha)\Big)=\nonumber\\
&=\dfrac{e^{\gamma}}{4\pi}\left(\tensor{F}{_a_c}\tensor{F}{_b^c}-\dfrac{1}{4}\tensor{g}{_a_b}\FF\right)\ ,
\end{align}
where we have used two auxiliary results
\begin{align}
{\hdg\tensor{F}{_a_c}}{\hdg\tensor{F}{_b^c}}-\tensor{F}{_a_c}\tensor{F}{_b^c}=-\dfrac{1}{2}\FF \tensor{g}{_a_b}\ , 
\\ \tensor{F}{_a_c}{\hdg\tensor{F}{_b^c}}={\hdg\tensor{F}{_a_c}}\tensor{F}{_b^c}=\dfrac{1}{4}\GG \tensor{g}{_a_b}\ .
\end{align}
Thus, the spacetime metric stays the same provided that the charge $Q$ is replaced as $Q^2\rightarrow \widetilde{Q}^2+\widetilde{P}^2$. Following the described procedure, we can easily extend the electrically charged multi BH solutions derived in the previous sections to dyonic configurations. 
The transformed electromagnetic 2-form $\widetilde{\textbf{F}}$ can be obtained from the previously introduced gauge potential $\pf{A}=-\chi(\rho, z)\df t$, 
\begin{align}
\widetilde{\textbf{F}}&=\text{sin}\alpha e^{\gamma}\rho e^{-2U(\rho, z)}(-(\partial_z\chi)\df\rho\wedge\df\phi+(\partial_\rho\chi)\df z\wedge\df\phi)-\nonumber\\&-\text{cos}\alpha((\partial_\rho\chi) \df\rho\wedge\df t+(\partial_z\chi)\df z\wedge\df t) \ ,
\end{align}
while the metric tensor is unchanged up to the trivial aforementioned charge redefinition. 
Notice that for $\alpha=\pi/2$ one recovers purely magnetic solutions. 

To give an explicit example, we may apply the duality transformation to a simple case of an electrically charged ModMax BH. For an arbitrary angle $\alpha$, the seed (\ref{eq:1MM}) is mapped to a dyonic solution. In canonical spherical coordinates, it is described by the following electromagnetic field
\begin{align}\label{eq:Fdyon}
\widetilde{\textbf{F}}&=-\dfrac{Qe^{-\gamma}}{r^2}\text{cos}\alpha\,\df t\wedge\df r+Q\text{sin}\alpha\, \text{sin}\theta\,\df\theta\wedge\df\phi =\nonumber\\
&=-\dfrac{\widetilde{Q}e^{-\gamma}}{r^2}\df t\wedge\df r+\widetilde{P}\text{sin}\theta\,\df\theta\wedge\df\phi\  ,
\end{align}
and a metric function
\begin{align}
f(r)=1-\dfrac{2M}{r}+\dfrac{(\widetilde{Q}^2+\widetilde{P}^2)e^{-\gamma}}{r^2}\ .
\end{align}
The rotation parameter $\alpha$ in (\ref{eq:Fdyon}) is absorbed in the definition of new charges. 

\section{Discussion}

To the best of our knowledge, the presented class of Einstein-ModMax spacetimes can be regarded as the first example of an exact multiple BH solution in an NLE theory.  The derivation of the electrically charged solution relies on the generalised Weyl formalism, while the duality symmetry of the theory guarantees a straightforward extension to the magnetic, and more generally, dyonic configurations. In the nonextremal case, the electromagnetic interaction is insufficient to counterbalance gravitational attraction, resulting in the appearance of struts between BHs. After taking the extremal limit, conical singularities are smoothed out, which proves that the balancing mechanism is not only a feature of Maxwell's electrodynamics but can be present in its nonlinear modifications. To reach this equilibrium state, ModMax BHs have to be endowed with more charge per unit mass than the Reissner-Nordstr\"{o}m solutions. Such behaviour is a consequence of the charge screening, represented by the dimensionless factor $\gamma$. We have also shown that an example of a nonstatic ModMax multi-BH configuration can be easily obtained by performing minimal alterations to the original cosmological Majumdar-Papapetrou solution \cite{KT93}. 

Possible future lines of research are branching into two directions: enriching the structure of multi-ModMax BH spacetimes or constructing other types of solutions within ModMax electrodynamics. For instance, the solutions derived in the paper could be immersed in the Melvin-type universe, with the external magnetic field providing a balancing effect. When applied to the axisymmetric, stationary, electrovacuum spacetimes, a specific transformation of the complex Ernst potentials generates solutions with the added NUT charge \cite{Astorino2020}. Starting from the Majumdar-Papapetrou-ModMax BHs, a similar procedure might be used to map it to its NUT generalisation. A further, highly nontrivial endeavour is the construction of the rotating ModMax BH. The Newman-Janis procedure, utilised to build rotating solutions from the static ones in Maxwell theory, generally does not solve the Einstein-NLE equations \cite{KTS22}. It remains an open question whether its modification may resolve this issue or if a completely novel approach is required.

\begin{acknowledgments}
This work is supported by the Center for Research and Development in Mathematics and Applications (CIDMA) through the Portuguese Foundation for Science and Technology (FCT -- Fundaç\~ao para a Ci\^encia e a Tecnologia) under the Multi-Annual Financing Program for R\&D Units, PTDC/FIS-AST/3041/2020 (\url{http://doi.org/10.54499/PTDC/FIS-AST/3041/2020}),  2022.04560.PTDC (\url{https://doi.org/10.54499/2022.04560.PTDC}) and 2024.05617.CERN. This work has further been supported by the European Horizon Europe staff exchange (SE) programme HORIZON-MSCA-2021-SE-01 Grant No.\ NewFunFiCO-101086251. A. B. acknowledges support from the Croatian Science Foundation, Project No. IP-2020- 02-9614.
\end{acknowledgments}

\bibliographystyle{jhep} 
\bibliography{refsMM}
\end{document}